\documentstyle[preprint,aps]{revtex}

\title{\bf  
Multifragmentation through Exotic Shape Nuclei \\
in {\boldmath ${\alpha}$}(5GeV/u) + Au Reactions}

\author{
Tomoyuki Maruyama\\
Advanced Science Research Center,\\
Japan Atomic Energy Research Institute,\\
Tokai, Ibaraki 319-11, Japan }

\begin{document}

\maketitle

\begin{abstract}

We simulate the fragmentation processes in the {$\alpha$~$+$~Au~}
collisions at a bombarding energy of 5 GeV/u using 
the simplified RQMD approach plus the statistical decay model.
We find from the simulation that the angular-distribution of 
the intermediate mass fragments has a sideward peak,  
more strongly in the transverse direction than in the beam-direction,  
when the intermediate nucleus formed by the dynamical process has an annular 
eclipse shape, which explains the experimental results. 

\end{abstract}

\newpage

Multifragmentation has been attracting attention as one of the most important 
aspects of light- and heavy- ion reactions in the intermediate and high energy
region \cite{expfr,Botvina}.
It is speculated that the decay of a highly excited nuclear system
carries the information about the nuclear equation of state (EOS) and
the liquid-gas phase transition of low density nuclear matter.

The very interesting results have been reported 
by the KEK experimental group
for the angular-distribution of the intermediate mass fragments (IMF)
in proton (12 GeV) and alpha (5 GeV/u) induced reactions \cite{KEK}.
There are two components in the experimental data for the angular-distribution
of IMF: one is forward peaking, and the other has a sideward peak
at $\theta_{{\rm lab}} = 70^\circ$. 
Recently this group has made new experiments of the proton induced reactions 
to choose only the central events using two coincidenct  IMFs 
with the opposite azimuthal direction on the $Au$ target \cite{KEK2}.
The component with the forward peak disappear, and the sideward
peak is enhanced very much and can be seen clearly even 
in the logarithmic scale \cite{KEK2}.

In this reaction we consider that some high energy pions, 
protons and light fragments such as deuteron are emitted forwards 
immediately after starting the collision, followed by 
IMFs created through the thermal decay of a hot intermediate nucleus.
If the hot intermediate nucleus decays isotropically, the angular-distribution
of IMF should have a forward peak.
Hence the sideward peak should be explained by the hypothesis
that the intermediate nucleus has an exotic shape and  expands 
more strongly in the transverse direction than in the beam-direction.

In the past H\"uffner and Sommermann \cite{Hueffner} suggested 
the trumpet-shaped hole to explain the enhanced backward emission 
of heavy fragments in high energy proton-induced reactions.
In recent years, moreover, the decay from the non-spherical nuclei has been  
suggested for the multifragmentation in heavy-ion collisions 
\cite{Moretto,Bauer,Bauer2,Xu}.

In this letter we theoretically analyze the fragmentation processes
of ${\alpha}$(5GeV/u)$~+~Au$ reaction,
and discuss the relation between the angular-distribution
and the shape of the intermediate nucleus.
We choose only the alpha- induced reaction here since 
at the bombarding energies of 12 GeV the elementary processes of 
two-body collisions are too complicated for many open channels.
Furthermore our study is restricted only to 
the central events for the reason mentioned above.
 
Our purpose is to clarify the fragmentation mechanism qualitatively, 
particularly in view of the shape of the intermediate nucleus.
For this purpose we should use a dynamical model which automatically givse 
the nuclear phase-space distribution at the intermediate time stage, 
which has to be assumed in statistical calculations 
\cite{Botvina,Hueffner,Bauer2}.
The Quantum Molecular Dynamics (QMD) approach \cite{Aich} is commonly used 
as a dynamical model for the theoretical study of fragmentation.
In this approach baryons are described as Gaussian wave packets, and their
dynamical motions are given by a mean-field and two-body collisions. 
Toshiki Maruyama et al. \cite{Toshi} and T. C. Sangster et al.\cite{Frank} 
have succeeded to reproduce experimental data of fragment multiplicities in 
heavy-ion collisions around several 10 MeV/u by using QMD together  
with the statistical decay model \cite{SDecay,SMM}.

At relativistic energies the Lorentz covariant transport approach is
desired because all nuclei and fragments must hold 
the consistent phase-space distribution under the Lorentz transformation.
In fact, these relativistic effects clearly appear in the multiplicity 
of alpha particles in the heavy-ion collisions 
even at $E_{\rm lab} \sim$ 1 GeV/u \cite{TOMO2}.
Therefore 
the Relativistic QMD (RQMD) approach \cite{Sorge,MARU1} should be the most
useful theoretical model for the present purpose.

In order to reduce the computation time,
we use the simplified version of RQMD (RQMD/S) \cite{Reltf},
where we take the time fixations to equalize all time coordinates 
of particles in the reference frame.
This new definition still hold the Lorentz covariance in the mean-field.
In Ref. \cite{Reltf}, then, we have confirmed that the RQMD/S give almost 
the same results as the full RQMD up to 6 GeV/u for the transverse flow,
which is thought to be the most sensitive observables to relativistic 
effects at present \cite{MARU2}. 
In the treatment of the two-body collisions, furthermore, 
we use the prescription of Ref. \cite{Kodama} to keep the Lorentz covariance
in the collisions within our time-fixation scheme around several GeV/u energy 
region \cite{Kodama,GyWolf}. 

Now we investigate the origin of the experimental results for the IMF 
angular-distribution by simulating the dynamical stage of 
{$\alpha$(5 GeV/u)~$+$~Au~} collisions with RQMD/S.  
The actual calculations are made in the following way.
First the initial distribution at rest is generated by the cooling 
method \cite{Cooling} and boosted according to the bombarding energy.
Second we perform the RQMD/S calculations and obtain the dynamical fragment 
distribution.
Third we boost each dynamical fragment to its rest frame and evaluate 
its excitation energy.
Finally we calculate the statistical decay \cite{SDecay} from 
the dynamical fragments and obtain the final fragment distribution.  

We use a Skyrme-type interaction with 'hard' EOS 
(the incompressibility $K$ = 380 MeV) parameterized 
in Refs. \cite{Aich,Bert} for the effective interactions. 
In addition, the symmetry force and the Coulomb force is introduced 
to get a correct isospin of a fragment in the simulation.
The Lorentz scalar Coulomb force can give correct effects to particle motions
in a relatively low energy region inside the fast moving matter.

For the cross-section of two baryon collisions we use the Cugnon's 
parameterization \cite{Cug} for an elastic channel and 
the Wolf's formulation \cite{GyWolf,Giessen} for inelastic channels including three baryonic 
resonances: $\Delta, N^{*}(1440)$ and $N^{*}(1535)$. 
These resonances can decay into nucleons and mesons ($\pi$ and $\eta$) 
\cite{Giessen}.
As for the parameters of the inelastic channels we use the values used 
in Ref. \cite{Niita}, since the parameters in Ref. \cite{GyWolf,Giessen} 
give unphysical large 
cross sections above {$E_{\rm lab}$ = 1.5 GeV \cite{Niita}.

In the fragmentation process the mean-field at low density
is considered to play an important role \cite{Moretto}.
In order to simulate the low density behavior we use two kinds 
of the width parameter $L$ for the Gaussian wave packets.
The first (case I) is defined by $L = 0.92$ fm given by Ref \cite{Toshi}, and
the second (case II) by $L = 0.625$ fm.
The Gaussian width does not affect bulk properties of ground states 
made by the cooling method \cite{Cooling}.
In the dynamical process, however, the difference of these two cases 
should describe the different instability in the low density region:
in the dilute medium the attractive force between nucleons are stronger 
in the case I than in the case II.

We show  the baryon and pion distributions in the coordinate space with a
12 fm/c time step, projecting on the $xz$- plane, restricted with positions 
$|y| < 1$ fm (upper columns), and the $xy$- plane, restricted with positions 
$|z| < 1$ fm (lower column), in Fig. 1 (case I) and in Fig. 2 (case II).
There $z$-axis and $xy$-plane are defined as 
the beam-direction and the reaction plane, respectively. 
These figures include results of twenty simulations for an impact-parameter 
$b = 0$ fm, and the blue, red, and yellow circles denote 
the nucleons, baryonic resonances and mesons, respectively.

Around the time step $t = 16$ fm/c a lot of resonances and 
pions are produced and propagate forwards in both cases.
After $t = 28$ fm/c, these high energy pions and nucleons are emitted 
forwards. 
At this step, the shapes of the intermediate nuclei are different between 
the two cases. 
In Fig. 2 (case II), the empty region appears in the center;
namely the intermediate nucleus 
with the annular eclipse shape is constructed through the reaction.
After that this exotic intermediate nucleus slowly expands sidewards, and 
disintegrates into some fragments.  
Apparently this fragmentation process is the multifragmentation.
On the other hand, in Fig. 1 (case I), the whole nucleons expand 
almost isotropically emitting the nucleons and finally the central part 
is shrunken again and form one big fragment with the small forwards velocity. 
Please note that the shape of the intermediate nucleus in the case II 
is similar to that in Ref. \cite{Bauer}.
However both formation processes are quite different, and our process does not
produce so clear ring/doughnut shape.

Next we evaluate the angular-distribution of fragments in these two cases.
Events are restricted to the impact-parameter $b < 3$ fm 
because the main contribution to the sideward peak component
is to come from central collisions.
In the actual calculation we perform six hundred QMD simulations and 
one hundred cascade calculations of the statistical decay model \cite{SDecay}
for each QMD simulation.
The numerical errors are found to be negligibly small.

In Fig. 3, we show the angular-distribution of two kinds of 
fragments with charge $Z = 1$ (open circles) and
$3 \leq Z \leq 20$ (IMF) (full squares), for case I in Fig. 3a
and for case II in Fig. 3b.
In this figure we do not show a comparison with experimental data \cite{KEK} 
because  the experimental group has given only data for fragments 
with $Z = 9$,
and we do not have enough computer power to get sufficient statistics for 
individual fragments.

In case I, both the angular-distributions of the two kinds of fragments 
have a forward peak. 
In case II, however, the IMF angular-distribution has a 
sideward peak and this result agrees with the experiment qualitatively.
In addition we can see that  
more IMFs are generated in case II than in case I.
We have found in the simulations that this sideward peak of the 
IMF angular-distribution is very much correlated 
with the shape of the intermediate nucleus, i.e. the annular eclipse shape.

The difference of the two processes (case I and II) must be caused by 
properties in the low density region.
Immediately after the incident alpha drills a hole in the $Au$ target,
the baryon-baryon collision process produces the side-directed force to
the medium, and the surface along this hole must be very steep.
The large Gaussian width in case I, however, makes a gentle surface, and 
diffuse the density into the hole region.
The nucleons along the hole, feeling a rather strong attractive mean-field, 
are gathered, and form a large compound nucleus. 
The IMFs are created through the evaporation and the binary fission.

In case II, on the other hand, the small width weakens 
the attractive mean-field in the low density region, and
the transverse expansion becomes stronger than this attractive force.
This expansion dilutes the medium, and increases the instability. 
Then the medium must be vaporized, and multifragmentation occurs. 
>From this consideration we can understand the reason of the larger 
cross-section in case II (see Fig. 3) as well as the sideward peak.

Finally we would like to comment on the contribution from non-central events.
With increase of the impact parameter the peak angle moves forwards.
In the middle impact-parameter region the intermediate nucleus
has a partial eclipse and expands sidewards.
Such a process is  also responsible for  the side-directed peak of the IMF 
angular-distribution while the peak angle depends on the impact-parameter.
However we cannot suppose that two IMFs generated through the partial eclipse
shape are observed in the azimuthally opposite directions the IMF-coincidence 
experiments \cite{KEK2}.

On the other hand we can consider the situation that the intermediate nucleus 
does not have a exotic shape, but has a very large angular-momentum.
In this case  the intermediate nucleus does not decay isotropically,
and two IMFs move to the azimuthally opposite directions.
However the rotational axis must direct almost perpendicularly to 
the reaction plane, and then the rotation affects the distribution of 
the azimuthal angle, but not that of the polar angle.
Thus the rotation of the intermediate nucleus cannot explain the sideward peak
in the first inclusive experiments \cite{KEK}.

To summarize, we have performed the QMD simulation for the 
$\alpha$ + Au collision at 5 GeV/u.
The dynamical instability in the dilute nuclear medium is simulated 
by changing the value of the Gaussian width parameters.
Then we conclude that this reaction constructs a hot nuclear system 
with the annular eclipse shape in the central collision.
This exotic intermediate nucleus expands sidewards 
and causes multifragmentation.
As a result of this process the IMF angular-distribution has a side-directed
peak.
This confirms findings in experimental results \cite{KEK,KEK2} 
showing clear evidence of multifragmentation.

\bigskip

The author would like to thank Prof. H. Horiuchi, Drs. A. Iwamoto,K. Niita, 
K.H. Tanaka, T. Murakami, Toshiki Maruyama and Mr. H. Ochiishi
for the useful discussions.
Dr. A. Ohnishi and Mr. Y. Nara give the initial distribution for the RQMD/S
simulations. 
This work is financially supported in part by the RCNP, Osaka 
University, as a RCNP Computational Nuclear Physics Project 
(Project No~94-B-04).


\newpage

\noindent
{\Large Figure Caption}

\bigskip 

\begin{itemize}
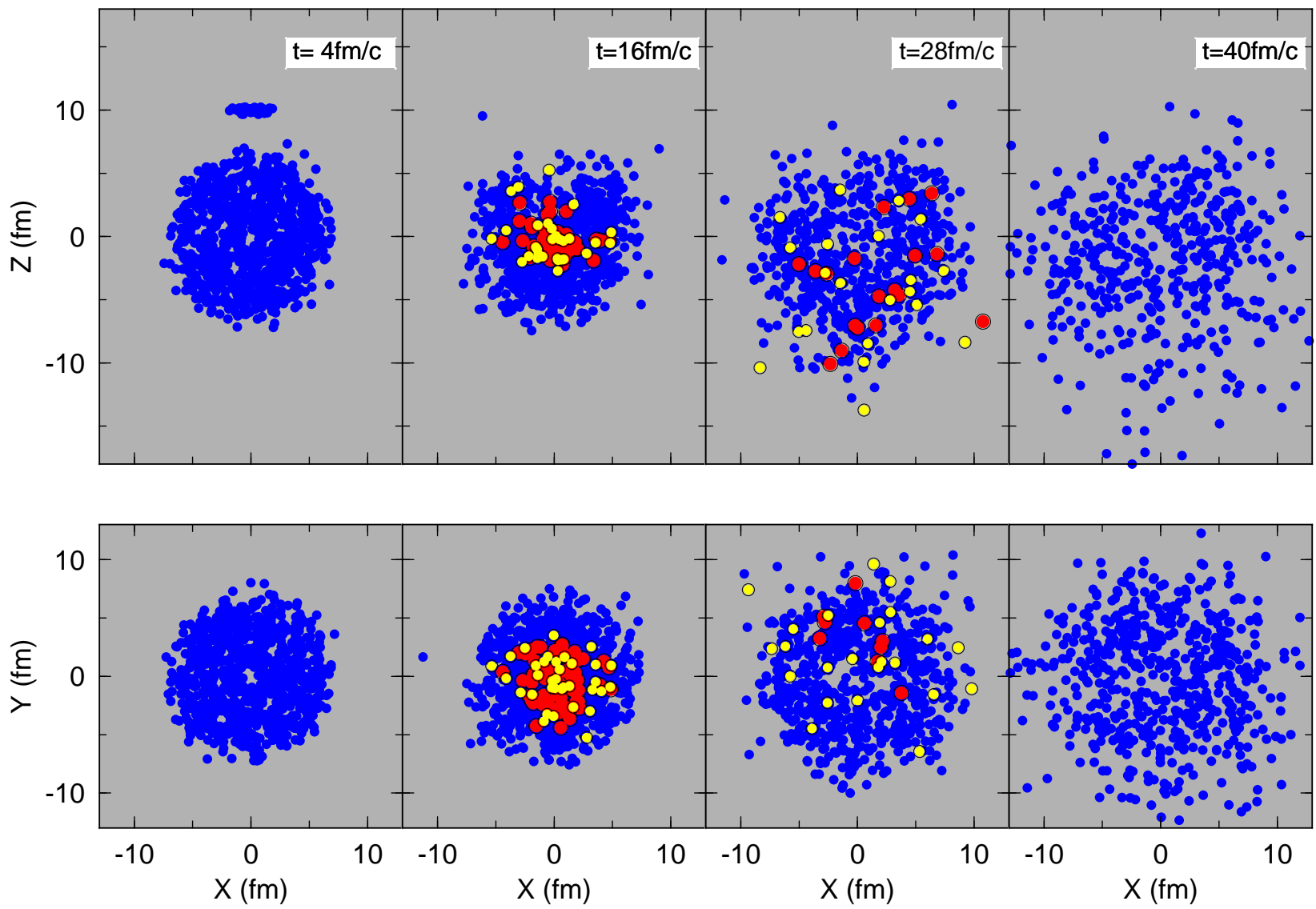
\item[Fig.1]
The time evolution of the baryon and meson distributions 
in the coordinate space 
at time steps  4, 16, 28 and 40 fm/c in {$\alpha$(5 GeV/u)~$+$~Au~}
collisions for the impact-parameter $b = 0$ fm in case I.
The upper columns show the distributions on $xz$- plane, restricted as 
$|y| < 1$ fm, 
while the lower columns on $xy$- plane, restricted as $|z| < 1$ fm. 
The blue, red, and yellow circles denote 
the nucleons, resonances and mesons, respectively.

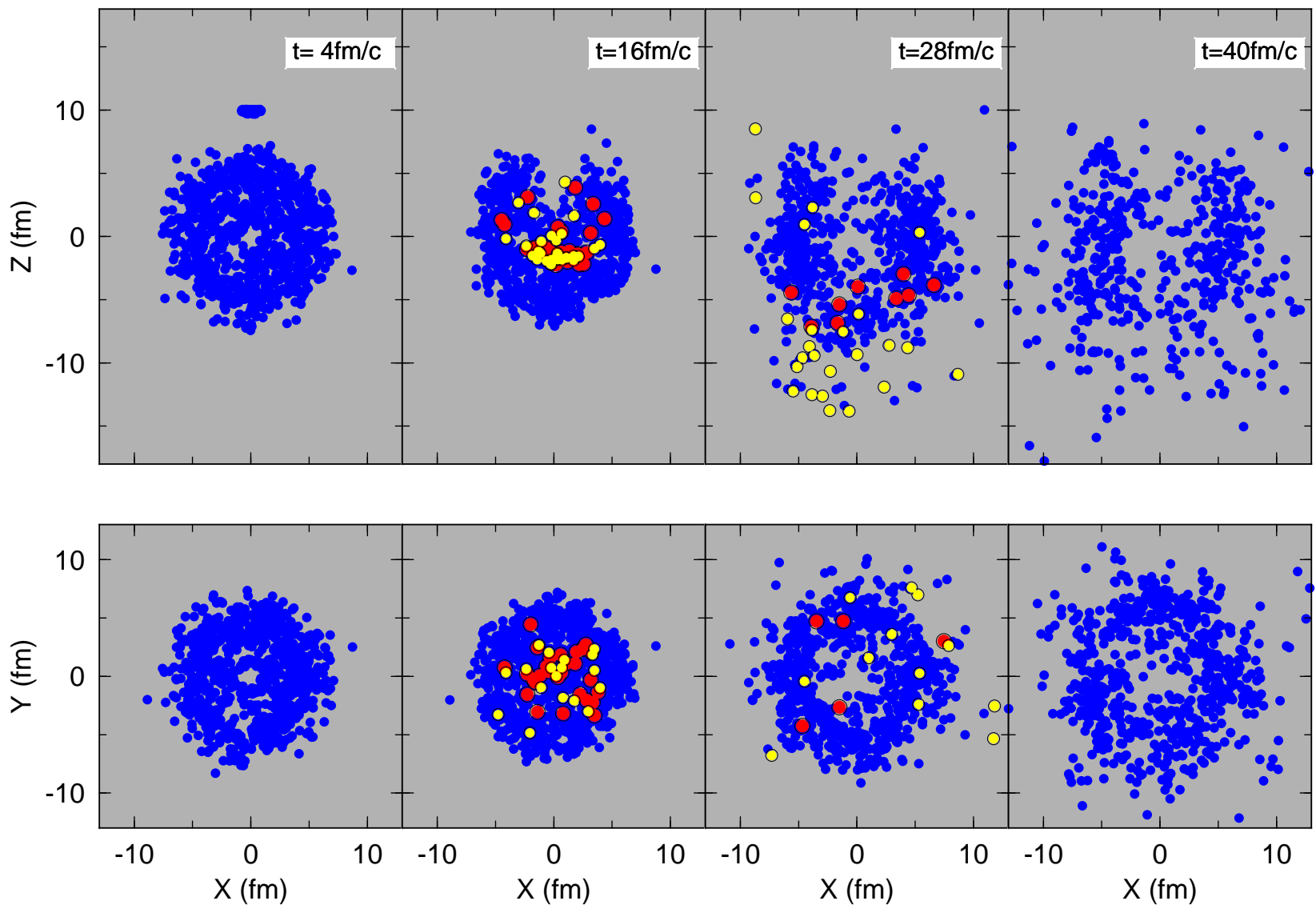
\item[Fig.2]
The same figure as in the Fig. 1, but for case II.

\item[Fig.3]
The angular-distributions of  fragments for case I (a) and case II (b):
$Z = 1$ (open circles) and $3 \leq Z \leq 20$ (full squares).
The cross-section for the second fragment is multiplied by 10.
Events are restricted with  the impact-parameter $b < 3$ fm.

\end{itemize}


\begin{thebibliography}{99}

\bibitem{expfr} 
L.G. Moretto and G.J. Wozniak, Ann. Rev. Nucl. Part. Science 
{\bf 43}, 379 (1993) and reference therein.
\bibitem{Botvina} 
A. S. Botvina et al., Nucl. Phys. {\bf A584}, 737 (1995), 
and references therein.
\bibitem{KEK} K.H. Tanaka et al., Nucl. Phys. {\bf A583}, 581 (1995).
\bibitem{KEK2} H. Ochiishi et al., KEK Preprint 95-37.
\bibitem{Hueffner} J. H\"uffner and H. M. Sommermann,
Phys. Rev. {\bf C27}, 2090 (1983).
\bibitem{Moretto} L. G. Moretto, K. Tso, N. Colonna and G. J. Wozniak, 
Phys. Rev. Lett. {\bf 69}, 1884 (1992).
\bibitem{Bauer} W. Bauer, G. F. Bertsch and H. H. Schultz, 
Phys. Rev. Lett. {\bf 69}, 1888 (1992).
\bibitem{Bauer2} L. Phair, W. Bauer, C. K. Gelbe, Phys. Lett. {\bf B314},
271 (1993).
\bibitem{Xu} H. M. Xu, J. B. Natowitz, C. A. Galigard
and R. E. Tribble, Phys. Rev. {\bf C48}, 933 (1993), and reference therein.
\bibitem{Aich} J. Aichelin, Phys. Rep. {\bf 202}, 233 (1991), and
reference therein.
\bibitem{Toshi} T. Maruyama, A. Ono, A. Ohnishi and H. Horiuchi,
Prog. Theor. Phys. {\bf 87}, 1367 (1992).
\bibitem{Frank} T. C. Sangster et al., Phys. Rev. {\bf 46}, 1409 (1992).
\bibitem{SDecay} F. P\"uhlhofer, Nucl. Phys. {\bf A280}, 267 (1977). 
\bibitem{SMM} 
J. S. Bondorf et al., Nucl. Phys. {\bf A443},321 (1985); {\bf A444},460
(1985);
A. S. Botvina et al., {\bf A507}, 649 (1990) and reference therein.
\bibitem{TOMO2} T. Maruyama, T. Maruyama and K. Niita, 
Phys. Lett. {\bf B358}, 34 (1995).
\bibitem{Sorge} H. Sorge, H. St\"ocker and W. Greiner, Ann. of
Phys. {\bf 192}, 266 (1989).
\bibitem{MARU1} T. Maruyama, S.W. Huang, N. Ohtsuka, G.Q. Li, A. F\"assler 
and J. Aichelin, Nucl. Phys. {\bf A534}, 720 (1991).
\bibitem{Reltf} T. Maruyama, K. Niita, T. Maruyama, S. Chiba, Y. Nakahara
and A. Iwamoto, nucl-th/9601010 (1996), submitted to Prog. Theo. Phys.
\bibitem{MARU2} T. Maruyama, G.Q. Li and A. F\"assler,
Phys. Lett. {\bf 268B}, 160 (1991); 
E. Lehmann, R.K. Puri, A. F\"assler, T. Maruyama,
G.Q. Li, N. Ohtsuka, S.W. Huang, D.T. Khoa, Y. Lotfy, M. A. Matin,
Prog. Part. Nucl. Phys. 30 (1992) 219
\bibitem{Kodama} T. Kodama, S. B. Duarte, K. C. Chung, R. Donangelo and
R. A. M. S. Nazareth, Phys. Rev. {\bf C29}, 2146 (1984).
\bibitem{GyWolf} Gy. Wolf, G. Batko, W. Cassing, U. Mosel,
K. Niita and M. Sch\"affer, Nucl. Phys. {\bf A517}, 615 (1990).
\bibitem{Cooling} T. Maruyama, A. Ohnishi and H. Horiuchi, Phys. Rev
{\bf C42}, 386 (1990). 
\bibitem{Bert} G. F. Bertsch and S. Das Gupta, 
Phys. Rep. {\bf 160}, 189 (1988).
\bibitem{Cug} J. Cugnon, T. Mizutani, J. Vandermeulen, Nucl. Phys.
{\bf A352}, 505 (1981). 
\bibitem{Giessen} Gy. Wolf, W. Cassing and U. Mosel,
Nucl. Phys. {\bf A545}, 139c (1992).
\bibitem{Niita} K. Niita, S. Chiba, T. Maruyama, T. Maruyama, H. Takada, 
T. Fukahori, Y. Nakahara and A. Iwamoto, Phys. Rev. {\bf C52}, 2620 (1995).
\end{thebibliography}
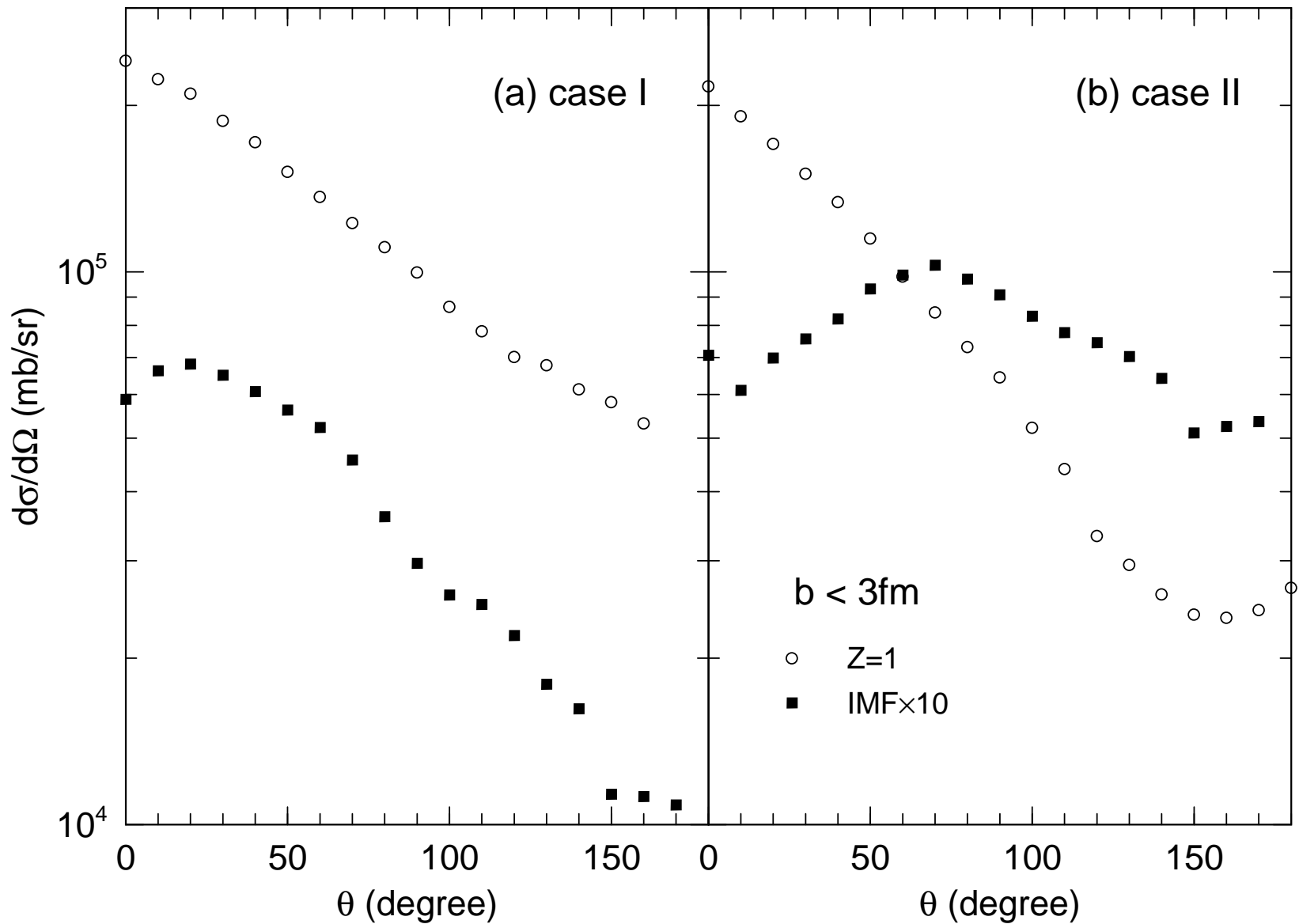
\end{document}